\documentstyle[twocolumn,seceq]{jpsj}

\def\runtitle{
BKT transition of Spin-1 XXZ Chains in a Staggered Magnetic Field
}
\def\runauthor{Masayoshi {\sc Tsukano} and Kiyohide {\sc Nomura}}

\title
{
Berezinskii-Kosterlitz-Thouless Transition of Spin-1 XXZ Chains \\
in a Staggered Magnetic Field
}

\author
{ 
Masayoshi {\sc Tsukano} and Kiyohide {\sc Nomura}$^1$
}

\inst
{
Institute for Solid State Physics,
University of Tokyo,Roppongi, Minato-ku, Tokyo 106

$^1$Department of Physics, Kyushu University,
Hakozaki, Higashi-ku, Fukuoka 812-81
}

\recdate
{
August 13, 1997
}

\abst
{
Phase diagram of $S=1$ XXZ chain in a staggered magnetic field is obtained
numerically. The Berezinskii-Kosterlitz-Thouless transition between the XY
and the antiferromagnetic phases is studied by the level-spectroscopy.
Moreover we find that there is no distinction between the Haldane and the
antiferromagnetic phases, since the Gaussian critical line does not appear,
in contrast to the $S=1/2$ case. It is expected that some hidden $Z_2\times Z_2$
symmetry breaking happens not only in the Haldane phase but also in the
antiferromagnetic one.
}

\kword
{
XXZ chain,
staggered magnetic field,
Berezinskii-Kosterlitz-Thouless Transition,
level-spectroscopy, hidden $Z_2\times Z_2$ symmetry
}

\begin{document}
\sloppy
\maketitle
\section{Introduction}
The groundstate properties of the spin-1 XXZ chain have been studied extensively
since Haldane's conjecture.\cite{Haldane83}
The XXZ chain is described by the following Hamiltonian,
\begin{equation}
H_{\rm XXZ}=\sum_{i=1}^L(S_i^xS_{i+1}^x+S_i^yS_{i+1}^y+\Delta S_i^zS_{i+1}^z),
\label{XXZ}
\end{equation}
where $\mbox{\boldmath $S$}_i^2=S(S+1)$.
For $S=1/2$ case, the groundstate was obtained exactly \cite{Yang73}
and its properties are understood well.\cite{Cloizeaux62,Luther75}
The ferromagnetic phase is for $\Delta<-1$ and the N\'eel phase for $\Delta>1$.
The XY phase for $-1<\Delta<1$ is characterized by the gapless excitation
and the power-law decay of the spin correlation function.
Before 1983, one had thought that the phase structure for $S=1/2$ is valid
for higher $S$, from the spin-wave picture.
However, Haldane \cite{Haldane83} predicted that
a novel phase $(\Delta_{c1}<\Delta<\Delta_{c2})$ appears between the XY phase
and the N\'eel phase for integer $S$, in which there is an excitation gap
and the spin correlation function decays exponentially,
in contrast to the half-odd integer $S$ case. He also suggested that
the Berezinskii-Kosterlitz-Thouless (BKT) transition
\cite{Berezinskii71,Kosterlitz74} occurs at $\Delta_{c1}$, and that
the transition at $\Delta_{c2}$ belongs to the 2D Ising universality class.
Nowadays his prediction is supported by many authors.
\cite{Affleck86,Schulz86,Nightingale86,Takahashi88,White93}
A number of numerical studies with respect to $S=1$ have shown that
$\Delta_{c1}=0 \sim 0.3$ and $\Delta_{c2}=1.18 \pm 0.02$.
\cite{Botet83,Kubo86,Gomez89,Nomura89b,Sakai90b,Yajima94}
Although it was difficult to estimate $\Delta_{c1}$ precisely due to the BKT
transition, recently it is settled that $\Delta_{c1}=0$.\cite{Kitazawa96a}
Kennedy and Tasaki \cite{Kennedy92} found that
the $S=$1 XXZ chain contains a hidden $Z_2\times Z_2$ symmetry,
and that the nature in the gapful phase, or Haldane phase,
is explained by the complete breaking of the $Z_2\times Z_2$ symmetry.

The hidden $Z_2\times Z_2$ symmetry of the groundstate can be controlled
by the bond-alternation, i.e. replacing the nearest neighbor interaction
in eq.(\ref{XXZ}) with $1-\delta (-1)^i$. In the isotropic case $\Delta=1$,
Affleck and Haldane \cite{Affleck85,Affleck87} argued that the bond-alternating 
system is mapped onto the O(3) nonlinear $\sigma$ model with the topological
term whose angle is $\theta=2\pi S(1-\delta)$, and that the excitation
becomes massless when $\theta/\pi$ is odd integer.
Therefore there should exist $2S$ transition points and $2S+1$
massive phases between $-1<\delta<1$. Oshikawa \cite{Oshikawa92}
suggested that, (although his argument is not applicable to half-odd
integer $S$) the $Z_2\times Z_2$ symmetry is essential for
the successive dimerization transitions.
The massive phases are classified into two types, depending on
whether the $Z_2\times Z_2$ symmetry is spontaneously broken or not,
and such two kinds of phases appear alternately.
For $S=1$, the Haldane-dimer transition occurs at
$\delta_c=0.2598$,\cite{Kato94,Yamamoto94,Totsuka95,Kitazawa96a}
and its universality is the same class as
the level $k=1$ SU(2) Wess-Zumino-Witten model.\cite{Totsuka95}
The whole phase diagram of the $S=1$ bond-alternating system with
arbitrary $\Delta$ was obtained by Kitazawa, Nomura and Okamoto.
\cite{Kitazawa96a} They pointed out that a Gaussian critical
line lies between the Haldane and the dimer phases, on the analogy
of the Ashkin-Teller model with the $Z_2\times Z_2$ symmetry.
Moreover they stated that the XY-Haldane phase boundary is just on
$\Delta=0$. The XY-Haldane and the XY-dimer transitions are of
the BKT type. The Haldane-N\'eel and the dimer-N\'eel transitions
belong to the 2D Ising universality class.

In this paper, as another example with the staggered interaction,
we investigate the BKT transition of the XXZ chain
in the staggered magnetic field, whose Hamiltonian is given by,
\begin{equation}
H=H_{\rm XXZ}+\lambda \sum_{i=1}^L(-1)^i S_i^z.
\label{XXZstag}
\end{equation}
The staggered magnetic field induces the antiferromagnetic (AF) order,
which is not the spontaneously symmetry breaking.
When $\lambda\gg1$, the AF phase characterized by the singlet groundstate,
AF long-range-order and gapful excitation should appear.
Recently, Alcaraz and Malvezzi \cite{Alcaraz95} examined this model for $S=1/2$.
They determined the XY-AF phase boundary by the phenomenological renormalization
group (PRG) method.\cite{Bonner83}
However, Okamoto and Nomura \cite{Okamoto96} criticized their study, since
the XY-AF phase transition is thought to be of the BKT type, and in that case
the simple application of the PRG method may lead to the false conclusion.
To overcome such a difficulty, one of the authers \cite{Nomura95} proposed
a new powerful method, "level-spectroscopy", to estimate the BKT transition 
point precisely. This idea is based on the SU(2)/$Z_2$ symmetry at the BKT
transition point. In this paper, we apply the level-spectroscopy to
the XY-AF transition in the staggered magnetic field, and we obtain
the phase diagrams for $S=1/2$ (in Fig. 1) and for $S=1$ (in Fig. 2).
In the $S=1/2$ case, there are two BKT lines and one Gaussian critical line
$(-1/\sqrt{2}<\Delta<1,\lambda=0)$. The Gaussian line can be determined by
the recent Kitazawa's method.\cite{Kitazawa96b}
The N\'eel phase $(\Delta>1,\lambda=0)$, in which the groundstate is doublet,
constructs a first-order transition line.  For $S=1$, on the other hand,
although the BKT line exists, any Gaussian line does not appear.
This is quite different from the bond-alternation case. The Haldane phase
$(0<\Delta<\Delta_{c2},\lambda=0)$ is connected continuously to the AF phase.
The N\'eel phase is on the line $\Delta>\Delta_{c2},\lambda=0$.
The existence of a hidden $Z_2\times Z_2$ symmetry is discussed in summary.
\section{Method and Symmetry}
We consider the 2D sine-Gordon model,
which is derived from the Hamiltonian (\ref{XXZstag}) by means of
the bosonization technique,\cite{Schulz86}
as an effective theory to describe the BKT transition.
The action is written by
\begin{equation}
S=\frac{1}{2\pi K}\int {\rm d}\tau {\rm d}x (\nabla\phi)^2
+\frac{y_\phi}{2\pi \alpha^2} \int {\rm d}\tau {\rm d}x \cos\sqrt{2}\phi,
\label{2DSG}
\end{equation}
where $\alpha$ is a short distance cut-off.
We also define the dual field $\theta(\tau,x)$ as
\begin{subequations}
\begin{eqnarray}
\frac{\partial}{\partial\tau} \phi(\tau,x) &=&
-{\rm i}K\frac{\partial}{\partial x} \theta(\tau,x), \\
\frac{\partial}{\partial x} \phi(\tau,x) &=&
{\rm i}K\frac{\partial}{\partial\tau} \theta(\tau,x).
\end{eqnarray}
\end{subequations}
Although the U(1) symmetry for the field $\phi$ is broken by
the second term of eq.(\ref{2DSG}), it remains in the dual field $\theta$.
Here we compactify $\phi,\theta$ on a circle with radius $1/\sqrt{2}$.
The winding number for a configuration of $\phi$ is nothing but
total magnetization along the $z$-axis.
The coupling constants $K$ and $y_\phi$ are renormalized
by the scaling transformation $\alpha \rightarrow \alpha {\rm e}^{{\rm d}l}$,
obeying the following renormalization group equations, \cite{Kosterlitz74}
\begin{subequations}
\begin{eqnarray}
\frac{{\rm d}}{{\rm d}l}y_0(l) &=& -y_\phi^2(l), \\
\frac{{\rm d}}{{\rm d}l}y_\phi(l) &=& -y_0(l)y_\phi(l),
\end{eqnarray}
\label{RGeqs}
\end{subequations}
where $y_0=K/2-2$. For the finite system, $l$ is related to $L$ by $l=\log L$.
The renormalization flow diagram is sketched in Fig. 3.
The perturbation of $\cos \sqrt{2} \phi$ is irrelevant for $K>4$
and the trajectory flows into the Gaussian fixed line $(y_0>0,y_\phi=0)$,
which corresponds to the XY phase with no symmetry breaking.
When $K<4$, $\cos \sqrt{2} \phi$ is a relevant operator
and $y_\phi$ goes to infinity.
In this case, $\langle\phi\rangle=\pi/\sqrt{2}$
and the translational symmetry is broken. This means that the groundstate has
AF long-range-order caused by the staggered magnetic field.
The BKT line $(y_\phi=y_0)$ is the boundary between the XY and the AF phases,
and $K$ is renormalized to 4.
The scaling dimension of the vertex operator
$O_{n,m}=\exp({\rm i}n\sqrt{2}\phi)\exp({\rm i}m\sqrt{2}\theta)$
on the Gaussian model $(y_\phi=0)$ is given by
\begin{equation}
x_{n,m}=\frac{1}{2}\left(n^2K+\frac{m^2}{K}\right).
\label{xnm}
\end{equation}
Thus the scaling dimension of $\cos\sqrt{2}\phi$ becomes 2
at the BKT fixed point $(y_\phi=y_0=0)$.
By the way, besides $\cos \sqrt{2} \phi$, there are four operators
which become marginal at the BKT fixed point. They are $\sin \sqrt{2} \phi$,
$\exp (\pm {\rm i}4\sqrt{2}\theta)$ and the marginal operator defined by
\begin{equation}
M=\frac{\alpha^2}{K}(\nabla\phi)^2.
\label{M}
\end{equation}
However, this degeneracy splits by the logarithmic corrections
on the BKT line away from the BKT fixed point $(y_\phi=y_0=0)$.
Next we consider how such splitting happens.
Here we note that $M$ and $\cos \sqrt{2} \phi$ are hybridized near the BKT line,
\begin{eqnarray}
A &=& M+\cos\sqrt{2}\phi, \\
B &=& \sqrt{2}\cos\sqrt{2}\phi-\frac{1}{\sqrt{2}}M.
\end{eqnarray}
The coefficients are determined by the orthogonal condition
$\langle A(\tau_1,x_1)B(\tau_2,x_2)\rangle=0$.
The renormalized scaling dimensions of the operators,
$A, B, \sin \sqrt{2}\phi$ and $\exp(\pm {\rm i}4\sqrt{2}\theta)$ are given as,
up to the lowest order in $y_0$,
\begin{subequations}
\begin{eqnarray}
x_0(l) &=& 2-y_0(l)\left(1+\frac{4}{3}t\right),
\label{x0} \\
x_1(l) &=& 2+2y_0(l)\left(1+\frac{2}{3}t\right),
\label{x1} \\
x_2(l) &=& 2+y_0(l),
\label{x2} \\
x_3(l) &=& 2-y_0(l),
\label{x3}
\end{eqnarray}
\label{x}
\end{subequations}
respectively.\cite{Nomura95}
An additional parameter $t$ plays the role of the deviation from the BKT line,
$y_\phi/y_0=1+t$, and $y_0(l)=l^{-1}=(\log L)^{-1}$ when $t\ll1$.
As an application of the conformal field theory (CFT), it is known that
the scaling dimension is related to the finite-size gap of the periodic system,
\cite{Cardy84}
\begin{equation}
E_n(L)-E_g(L)=\frac{2\pi v}{L}x_n,
\label{CFT}
\end{equation}
where $v$ denotes the spin-wave velocity.
Hence the eigenvalues corresponding to $x_0$ and $x_3$
cross linearly on $t$ and degenerate on the BKT line, reflecting the SU(2)/$Z_2$
symmetry of the BKT line.

We apply the above sine-Gordon theory to the original spin system.
The Hamiltonian (\ref{XXZstag}) is invariant under spin rotation
around the $z$-axis, translation by two sites
$(\mbox{\boldmath $S$}_i\rightarrow \mbox{\boldmath $S$}_{i+2})$ and
space inversion $(\mbox{\boldmath $S$}_i\rightarrow\mbox{\boldmath $S$}_{L-i};
\mbox{reflection on a spin site})$.
Therefore eigenstates are classified by total spin moment
$(S_T^z=\sum_{i=1}^LS_i^z)$, wave number $(q=2\pi n/L)$ and parity $(P=\pm1)$.
Besides them, symmetry operation includes modified time reversal 
$(T:S_i^z\rightarrow-S_{i+1}^z, S_i^\pm\rightarrow-S_{i+1}^\mp)$.
The symmetry of the sine-Gordon operators is summarized in  Table \ref{table1}.
As a result, we can determine the BKT line by
level-crossing of the excitation with $q=0,S_T^z=0,P=1,T=1$
and that with $q=0,S_T^z=4,P=1$.
\begin{table}
\caption{
Correspondence of the excitations to the sine-Gordon operators.
}
\label{table1}
\begin{tabular}{@{\hspace{\tabcolsep}\extracolsep{\fill}}ccc}
\hline
Symmetry & Operator on & Notation of \\
$(q,S_T^z,P,T)$ & sine-Gordon model & scaling dimension \\ \hline
$(0,0,+,+)$ & $M$ & $x_0$ \\
$(0,0,+,+)$ & $\cos \sqrt{2}\phi$ & $x_1$ \\
$(0,0,-,-)$ & $\sin \sqrt{2}\phi$ & $x_2$ \\
$(0,4,+,\ \ \,)$ & $\exp(-{\rm i}4 \sqrt{2} \theta)$ & $x_3$ \\
\hline
\end{tabular}
\end{table}
\section{Numerical Results}
The phase diagram of the $S=1$ system is shown in Fig. 2.
We demonstrate numerically the XY-AF transition, fixed $\Delta=-0.5$ especially.
Figure 4 shows some low-energy excitations of the finite-size system with
$L=14$.
The critical point $\lambda_c$ is determined by the crossing point of
the excitation energy in $q=0,S_T^z=0,P=1,T=1$ subspace
and that in $q=0,S_T^z=4,P=1$ subspace.
We obtain $\lambda_c(L)$ up to $L=18$ (see Fig. 5), and then
we estimate that $\lambda_c=0.855\pm0.001$ in the thermodynamic limit
$L\rightarrow\infty$. The correction of $1/L^2$ originates from
the irrelevant field with the scaling dimension 4.
At the critical point $\Delta=-0.5, \lambda=0.855$ obtained above,
we also calculate the averaged scaling dimensions defined as
$(x_0+x_2)/2$ and $(2x_0+x_1)/3$, which eliminate the logarithmic correction,
from the energy gaps in eqs.(\ref{x}),(\ref{CFT}).
As is shown in Fig. 6, they  converge to 2 within $1\%$ error.
The conformal anomaly is also estimated to be $c=0.993$.
These features confirm that the XY-AF transition is of the BKT type.
The XY-AF transition line approaches the point $\Delta=\lambda=0$,
and close to $\Delta=\lambda=0$, it fits on $\lambda^2\propto-\Delta$ 
very well (see inset in Fig. 2).

To investigate the Gaussian critical line, we make use of the Kitazawa's method.
\cite{Kitazawa96b} According to him, when antiperiodic boundary condition
$(S_L^{x,y}=-S_0^{x,y},S_L^z=S_0^z)$ is taken, two low-lying energies with
$S_T^z=0$ have to degenerate at a Gaussian critical point.
For $S=1/2$, we calculate the energies of the antiperiodic system with
$L=20$ and $\Delta=0.5$. As is shown in Fig. 7, they cross linearly at
$\lambda=0$. This means that the Gaussian line is drawn on $\lambda=0$ and
is governed by the $c=1$ U(1) CFT.\cite{Hamer86,Alcaraz87,Woynarovich87,Hamer88}
As to $S=1$, we perform the similar calculation about $L=14,\Delta=0.5$.
Unlike the $S=1/2$ case, the level-crossing is not observed in Fig. 8.
Thus the groundstate does not undergo the phase transition between
the Haldane and the AF phases.
\section{Summary and Discussion}
We have studied the effect of the staggered magnetic field on
the $S=1$ XXZ chain. The BKT transition line between the XY and the AF phases
has been obtained numerically by the level-spectroscopy.
Next we have tried to find the Gaussian critical line, using the Kitazawa's method. However, the Gaussian line is absent
and there is no distinction between the Haldane and the AF phases.
Then we suppose that the Hamiltonian has some hidden $Z_2\times Z_2$ symmetry
as the sine-Gordon model does, and we expect that the $Z_2\times Z_2$ symmetry
is spontaneously broken in both phases.
In contrast, the $S=1/2$ case can be interpreted more naturally from the hidden
$Z_2\times Z_2$ symmetry picture; it is a variant of the Ashkin-Tellar model.
One can consider that the hidden $Z_2\times Z_2$ symmetry is fully broken
for $\lambda<0$,unbroken for $\lambda>0$ and partially broken (N\'eel region)
on the line $\lambda=0,\Delta>1$. Note that the definition of the fully
$Z_2\times Z_2$ broken region and the unbroken region is artificial, similar to
the $S=1/2$ bond-alternating system where one can interchange the ordered and
the disordered region, taking the pair $(2i-1,2i)$  or $(2i,2i+1)$
in the string order parameter.\cite{Hida92,Kohmoto92,Takada92}

With respect to the $S=1$ case, the existence of the hidden $Z_2\times Z_2$
symmetry shall be more clarified by taking account of uniaxial single-ion
anisotropy, i.e. $D\sum_{i=1}^L(S_i^z)^2$, apart from the question
what is the explicit form of the $Z_2\times Z_2$ symmetry. At least when
$D\gg1$ and $\lambda=0$, the $Z_2\times Z_2$ symmetry breaking does not occur
because the groundstate is in the "large-$D$" phase.\cite{Kennedy92}
Thus, if the hidden $Z_2\times Z_2$ symmetry is contained in the Hamiltonian,
the Gaussian critical line has to emerge between the AF and the large-$D$
phases. We are now investigating these possibilities.\cite{Tsukano97}
\section*{Acknowledgements}
We would like to thank A.Kitazawa, M.Takahashi and M.Yajima
for stimulating discussions and helpful comments.
Most numerical calculations were done by the FACOM VPP500 of the supercomputer
Center, the Institute for Solid State Physics, the University of Tokyo.
This research is supported in part by Grants-in-Aid for Scientific Research
(C) No.09740308 from the Ministry of Education, Science and Culture, Japan.

\newpage
\section*{Figure Caption}
\begin{figure}

\caption{
Phase diagram of the $S=1/2$ system.
The XY-AF transition is of the BKT type,
and the dotted line denotes the Gaussian critical line.
The N\'eel phase (doubly degenerate) makes the first-order transition line
(bold line).
}
\label{fig1}

\caption{
Phase diagram of the S=1 system.
The inset shows the blow-up region for $-0.1\leq\Delta<0$
on double logarithmic scale.
We obtain that $\sigma=0.52\pm 0.02$ on assumption that
$\lambda\propto(-\Delta)^\sigma$ near $\Delta=\lambda=0$.
}
\label{fig2}

\caption{
Renormalization flow diagram of the 2D sine-Gordon model.
}
\label{fig3}

\caption{
Low-energy excitations as $L=14$ and $\Delta=-0.5$.
Eigenstates belong to $q=0,S_T^z=0,P=1,T=1(\circ)$,
$q=0,S_T^z=0,P=-1,T=-1(\diamond)$ and $q=0,S_T^z=4,P=1(+)$ subspace,
respectively.
}
\label{fig4}

\caption{
Estimation of the BKT transition point as $\Delta=-0.5$.
Using the system sizes $L=8\sim18$, we obtain that
$\lambda_c=0.855\pm 0.001$.
}
\label{fig5}

\caption{
Scaling dimensions, $(x_0+x_2)/2 (\circ)$ and $(2x_0+x_1)/3 (\diamond)$,
at the critical point $\Delta=-0.5, \lambda=0.855$.
}
\label{fig6}

\caption{
Low-lying energies in $S_T^z=0$ subspace of the $S=1/2$ system
with $L=20$ and $\Delta=0.5$, on antiperiodic boundary condition.
}
\label{fig7}

\caption{
Low-lying energies in $S_T^z=0$ subspace of the $S=1$ system
with $L=14$ and $\Delta=0.5$, on antiperiodic boundary condition.
}
\label{fig8}
\end{figure}
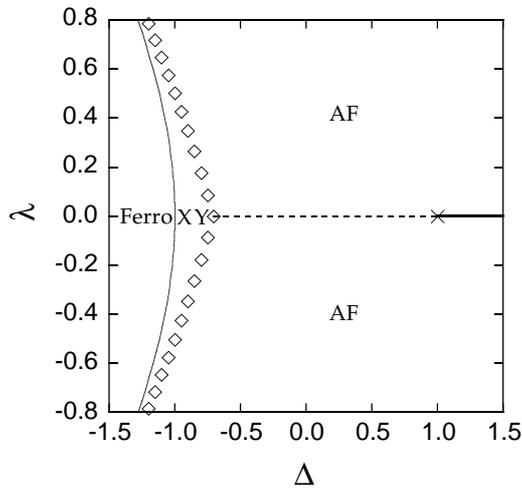
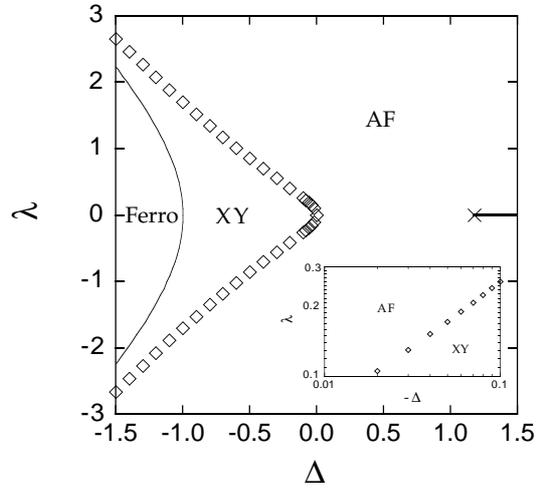
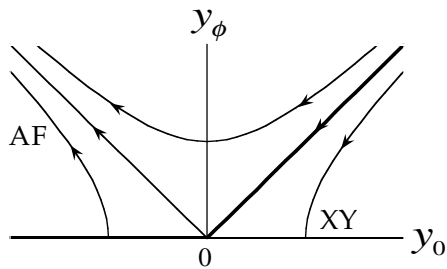
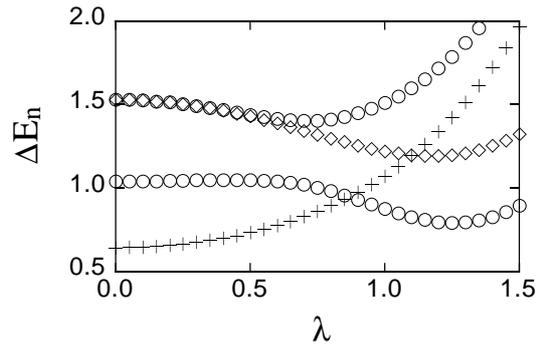
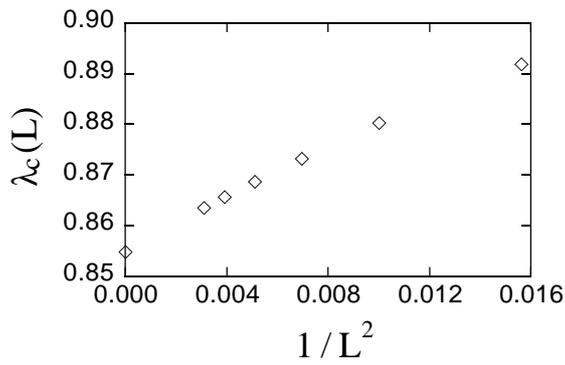
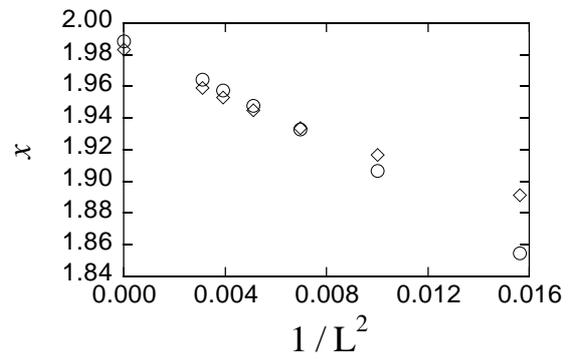
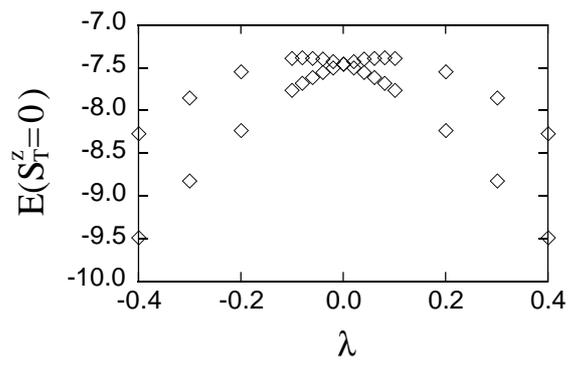
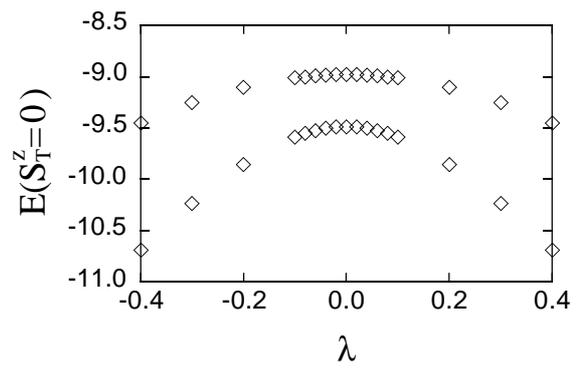
\end{document}